%% file: main.tex
\documentclass[10pt]{article}

\usepackage{algorithm}
\usepackage{graphicx}
\usepackage{textcomp}
\usepackage{xcolor}
\usepackage[utf8]{inputenc}
\usepackage{listings}
\usepackage{float}
\usepackage{subfigure}
\usepackage{caption}
\usepackage{url}
\usepackage{multirow}
\usepackage{commath}
\usepackage{algpseudocode}
\usepackage{etoolbox}
\usepackage{varwidth}
\usepackage{tikz}
\usetikzlibrary{shapes,arrows}
\usepackage{colortbl}
\usepackage{adjustbox}
\usepackage[margin=1in]{geometry}

\newcommand{\pluseq}{\mathrel{+}=}

\newcolumntype{P}[1]{>{\centering\arraybackslash}m{#1}}

\definecolor{codegreen}{rgb}{0,0.6,0}
\definecolor{codeblack}{rgb}{0.30,0.30,0.30}
\definecolor{codepurple}{rgb}{0.58,0,0.82}
\definecolor{backcolour}{rgb}{0.95,0.95,0.92}
 
\lstdefinestyle{codeListStyle}{
    backgroundcolor=\color{backcolour},
    commentstyle=\color{codegreen},
    keywordstyle=\color{magenta},
    numberstyle=\scriptsize\color{codeblack},
    stringstyle=\color{codepurple},
    breaklines=true,
    breakatwhitespace=true,
    numbers=left,
    xleftmargin=0.2in,
    xrightmargin=0.05in,
    basicstyle=\linespread{1.2}\ttfamily\footnotesize,
    numbersep=5pt,
    frame=single,                  
    tabsize=2
}

\AtBeginDocument{%
  \providecommand\BibTeX{{%
    \normalfont B\kern-0.5em{\scshape i\kern-0.25em b}\kern-0.8em\TeX}}}

\begin{document}

\title{LLVM Static Analysis for Program Characterization and Memory Reuse Profile Estimation}

\author{Atanu Barai$^{1,3}$,  Nandakishore Santhi$^2$, Abdur Razzak$^1$, Stephan Eidenbenz$^2$, \and Abdel-Hameed A. Badawy$^{1,2}$}

\date{
    \small
    ${}^\ast$ Klipsch School of ECE, New Mexico State University, Las Cruces, NM 80003, USA\\
    ${}^\ddagger$ Los Alamos National Laboratory, Los Alamos, NM 87545, USA\\
    \{atanu, arazzak\}@nmsu.edu, \{nsanthi, eidenben\}@lanl.gov, \{badawy\}@nmsu.edu\\
    $^3$ Intel Corporation, Folsom, CA 95630, USA\\
}

\maketitle

\begin{abstract}
\input{sections/abstract}
\end{abstract}


\section{Introduction}
\input{sections/indroduction}
\label{sec:sa:intro}

\begin{figure}[tbp]
    \centering
    \includegraphics[trim=35mm 40mm 35mm 40mm, clip, width=.5\textwidth]{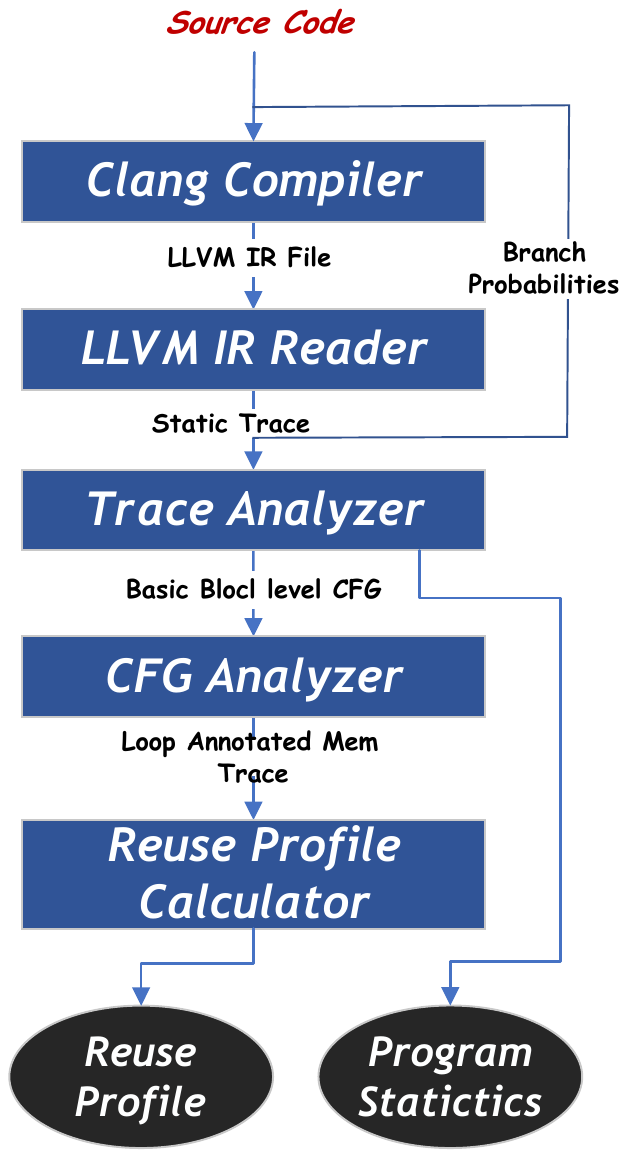}
    \vspace{-7mm}
    \caption{Overview of LLVM Static Analyzer}
    \vspace{-3mm}
    \label{fig:sa:overview}
\end{figure}

\section{Related Work}
\label{sec:sa: related_works}
\input{sections/relatedWorks}

\section{Methodology}
\label{subsec:sa:method}
\input{sections/methodology}

\begin{table}[h]
    \centering
    \caption{List of applications used to validate the static analyzer. LA and DM denote linear algebra and data mining kernels, respectively.}
    \vspace{-3mm}
    \resizebox{\linewidth}{!}{%
    \begin{tabular}{||c||P{5.0cm}||c||}
    \hline
    \textbf{App} & \textbf{Description} & \textbf{Domain}\\
    \hline\hline
    2mm & Two Matrix Multiplications & LA\\
    \hline
    atax & Matrix Transpose and Vector Multiplication & LA\\
    \hline
    bicg & BiCG Sub Kernel of BiCGStab Linear Solver & LA\\
    \hline
    doitgen & Multi-Resolution Analysis kernel & LA \\
    \hline
    mvt & Matrix Vector Product and Transpose & LA \\
    \hline
    gemver & Vector Multiplication and Matrix Addition & BLAS \\
    \hline
    gesummv & Scalar, Vector and Matrix Multiplication & BLAS \\
    \hline
    gemm & Matrix-Multiply C=alpha.A.B+beta.C & BLAS \\
    \hline
    symm & Symmetric Matrix-Multiply & BLAS \\
    \hline
    covariance & Covariance Computation & DM \\
    \hline
    \end{tabular}
    }
    \label{table:sa:benchmarks}
\end{table}

\section{Results}
\label{sa:results}
\input{sections/results}

\section{Limitations}
\label{sa:limitations}
\input{sections/limitations}

\section{Conclusion}
\label{sa:conclusion}
\input{sections/conclusion}

\section*{Acknowledgment}
This paper has been approved for public release with LA-UR-23-29319.
The authors would like to thank the anonymous reviewers for their feedback. 
This research is partially supported through the U.S. Department of Energy (DOE) National Nuclear Security Administration (NNSA) under Contract No. DEAC52-06NA25396, 
and by Triad National Security, LLC subcontract \#581326. 
Any opinions, findings, and/or conclusions expressed in this paper do not necessarily represent the views of the DOE or the U.S.\ Government.

\bibliographystyle{plainurl}
\bibliography{reference}

\end{document}

%% file: sections/abstract.tex
Profiling various application characteristics, including the number of different arithmetic operations performed, memory footprint, etc., dynamically is time- and space-consuming. On the other hand, static analysis methods, although fast, can be less accurate. This paper presents an LLVM-based probabilistic static analysis method that accurately predicts different program characteristics and estimates the reuse distance profile of a program by analyzing the LLVM IR file in constant time, regardless of program input size. We generate the basic-block-level control flow graph of the target application kernel and determine basic-block execution counts by solving the linear balance equation involving the adjacent basic blocks' transition probabilities. Finally, we represent the kernel memory accesses in a bracketed format and employ a recursive algorithm to calculate the reuse distance profile. The results show that our approach can predict application characteristics accurately compared to another LLVM-based dynamic code analysis tool, Byfl.

%% file: sections/indroduction.tex
With the help of advanced manufacturing and packaging techniques, complex CPUs contain billions of transistors condensed onto a single chip. Given their intricate nature, researchers must turn to performance modeling and program analysis tools to investigate the consequences of hardware design alterations, steering research in the correct direction. On the other hand, software developers can tune their software for particular hardware and conduct performance analysis using performance models. This approach also aids in selecting appropriate hardware for specific workloads. Consequently, program characterization and simulation tools facilitate hardware/software co-design.

A critical factor determining an application's performance is the number and types of instructions it executes. Since each instruction is usually executed many times during a program execution, the number of times each execution provides insight into the application's behavior. On the other hand, an application's performance largely depends on the data availability to the arithmetic logic units. The cache utilization of an application and its data access pattern determines its locality.

In analyzing a cache's performance, \textit{Reuse Distance Analysis}~\cite{Mattson:RD:IBM} is one of the commonly used techniques~\cite{performance:Sen:2013:ROM,locality:Zhong:2009:PLA}. Reuse distance is defined as the number of unique memory references between two references to the same memory location. When a memory location is accessed for the first time, its reuse distance is infinite. The reuse profile is the histogram of reuse distances of all the memory accesses of a program. For sequential programs, the reuse profile is architecture-independent, needs to be calculated only once, and can be used to quickly predict memory performance while varying the cache architectural parameters. Although researchers tried to speed up reuse distance calculation by parallelizing the algorithm~\cite{PARDA:Niu} and proposing analytical model and sampling techniques~\cite{chennupati:pmbs,ppt-sasmm}, it still requires collecting a large amount of memory trace which is both time and space consuming. Recently, researchers proposed several methods to statically calculate reuse profile from source code~\cite{rd_static_beyls,rd_static_matlab, performance:CaBetacaval:2003:ECM,rd_static_narayan}. However, their approaches require modifying the programs or are limited to a single loop nest with numeric loop bounds.

This paper\footnote{This work was part of the PhD dissertation of the first author~\cite{atanu-thesis}. Any opinions, findings, and/or conclusions expressed in this paper do not represent Intel Corporation’s views.} introduces an LLVM~\cite{Lattner:LLVM} static code analysis tool to gather program characteristics, including the number of different arithmetic operations, memory operations at both basic block and whole program level and the number of times each basic block is executed. It can also statically predict the reuse profile of the program without running the app or collecting the memory trace. It takes the source code and the branch probabilities as input and outputs different program statistics along with the reuse profile of the target kernel. Since we are interested in analyzing application kernels, our tool currently supports the reuse profile calculation of a single function. The prediction time of our model is input size invariant, making it highly scalable.

First, we dump the LLVM Intermediate Representation (IR) file using the \emph{Clang} compiler. Then, we deploy an LLVM IR reader to read the IR file and generate a static application trace. We parse the static trace using the trace analyzer and generate the program's Control Flow Graph (CFG). Each node in the CFG represents an LLVM basic block of the application. In this step, we also take the program branch probabilities for a specific input size to calculate the basic block execution counts and the number of different arithmetic/memory operations. We annotate CFG edges with branch probabilities and nodes with the execution counts and memory accesses. In the next step, we take the CFG as input, and using the branch probabilities and memory access information, we construct a loop annotated static memory trace of the program. Finally, we take the static memory trace and calculate the reuse profile using a recursive algorithm. We evaluate our analyzer with the program statistics collected from another LLVM binary instrumentation and code analysis tool, Byfl~\cite{Byfl}. The results show that the model accurately predicts program characteristics and reuse profiles, given the same LLVM version used for Byfl and our static analyzer.

%% file: sections/relatedWorks.tex
Researchers investigated different approaches of static code analysis~\cite{static_software_verification, static_software_testing, static_industrial_tool} and reuse profile calculation~\cite{rd_static_beyls,rd_static_matlab,performance:CaBetacaval:2003:ECM,rd_static_narayan}.

Cousot~\emph{et al.}~\cite{astree_comparison} described Astree, a static analyzer that provides runtime errors in C programs. However, they are bound to C program only, and their static analyzer cannot catch errors sometimes. Lindlan~\emph{et al.}~\cite{static_tool_obj_oriented_soft}, on the other hand, analyzes the Program Database Toolkit with both static and dynamic methods. Nevertheless, their static analysis needs an adequate front-end compilation. Bush~\emph{et al.}~\cite{bush_static_analyzer} describes a static compile-time analyzer that can detect various dynamic errors in some real-world programs. However, their static analysis approach is defined to some programs only.


Beyls~\emph{et al.}~\cite{Byfl} proposed a method that highlights code areas with high reuse distances and then tries to refactor the code to reduce the cache misses. However, they are required to modify the source code. Narayanan~\emph{et al.}~\cite{rd_static_narayan} proposed a static reuse distance-based memory model, but it is limited to loop-based programs. Cascaval~\emph{et al.}~\cite{performance:CaBetacaval:2003:ECM} presented a compile-time algorithm to compute reuse profile. However, they were also limited to loop-based programs and required program modification. Chauhan~\emph{et al.}~\cite{rd_static_matlab} statically calculated reuse distance only for Matlab programs.

In contrast, our approach is probabilistic and leverages the LLVM IR to compute the reuse profiles and other program metrics. As a result, it enables us to support a variety of programming languages supported by LLVM compiler infrastructure. It also generates accurate profiles without changing the source program, and the output is tested with state-of-the-art dynamic models.

%% file: sections/methodology.tex
In this section, we discuss our static analysis tool in detail. Figure~\ref{fig:sa:overview} shows different steps of the analyzer where it shows that we collect various program characteristics at an early stage of the analyzer. Further, we extend it to calculate the reuse profile to predict cache hit rates. In the following sections, we discuss these steps in detail. In the following sections, we will use the program in Figure~\ref{fig:sa:running_example} as a running example.

\subsection{Genarate LLVM IR File}
In the first step, we generate the LLVM IR of the target program using the `\textit{clang -g -c -emit-llvm example.c}' command. Here, command line options enable `source-level debug information generation,' `only preprocess, compile, and assemble steps,' and `LLVM IR file generation' options respective to their appearance. Thus, \emph{clang} only runs `preprocess' and `compilation' steps while generating the IR with source-level debug information.

\begin{figure}[tbp]
    \centering
    \lstset{style=codeListStyle}
    \begin{lstlisting}[language=C,numbers=left]
int main() {
    unsigned i, j, k, result, A = 10, arr[100][200];
    for (i = 0; i < 100; i++) {
        result += i;
        for (j = 0; j < 200; j++) {
            result = result * j + A;
            for (k = 0; k < 300; k++)
                result += k * k;
            arr[i][j] = result;
        }
    }
    return 0;
}
    \end{lstlisting}
    \caption{Running example program.}
    \label{fig:sa:running_example}
\end{figure}

\subsection{LLVM IR File Reader}
\label{subsubsec:sa:bcread}
The second step of the static analyzer is the LLVM Intermediate Representation (IR) reader written as an LLVM pass. The reader takes the IR and optionally function name as input and outputs a static trace of the target function. It is written as an \emph{LLVM pass} and uses LLVM compiler APIs to go over all the modules, functions, basic blocks, and instructions in the IR file.

Figure~\ref{fig:sa:static_trace} shows a portion of the trace generated by the IR reader for the example program. Here, under \emph{`Example'} module, there is \emph{`main'} function. It receives no argument and consists of several basic blocks (`e.g., entry, for.cond, for.body'). All the instructions under each basic block are also listed along with their position in the source \emph{`C'} program, operand list, and their types. For example, instruction \emph{`load'} in line 16 accesses \emph{`i'}, where \emph{`i'} is loaded from memory, compared against 100 (line 18), and upon comparison, a branch is taken (line 21 in the trace). Thus, we can determine the variables accessed by each memory operation (load/store) within a basic block. In the case of array references (if the reference results from GEP instruction), we determine the index variables. For \emph{`loops'}, we determine the loop control variable.

\begin{figure}[tbp]
    \centering
    \lstset{style=codeListStyle}
    \begin{lstlisting}[numbers=left][trim=10mm 20mm 10mm 20mm, clip, width=0.45\textwidth]
Module : Example.llvm
	Function <@ 0x7d8a88> : main : Definition : NonIntrinsic
		ArgList :
		BasicBlock <@ 0x7f3bd0> : entry
			Instruction <@ 0x7f3d38> : (0, 0) : alloca
				i32 : i32 1<@ 0x7f4390>
			............
			............
			Instruction <@ 0x7f7130> : (4, 12) : store
				i32 : i32 0<@ 0x7f4660>
				i32* : i<@ 0x7f43f8>
			Instruction <@ 0x7f7358> : (4, 10) : br
				label : for.cond<@ 0x7f7290>

		BasicBlock <@ 0x7f7290> : for.cond
			Instruction <@ 0x7f72f8> : (4, 17) : load
				i32* : i<@ 0x7f43f8>
			Instruction <@ 0x7f7680> : (4, 19) : icmp
				i32 : i32 %0<@ 0x7f72f8>
				i32 : i32 100<@ 0x7f7620>
			Instruction <@ 0x7f7a18> : (4, 5) : br
				i1 : cmp<@ 0x7f7680>
				label : for.end15<@ 0x7f78f0>
				label : for.body<@ 0x7f7810>

		BasicBlock <@ 0x7f7810> : for.body
			Instruction <@ 0x7f7cf8> : (6, 16) : load
				i32* : result<@ 0x7f4578>
        ............

    \end{lstlisting}
    \vspace{-5mm}
    \caption{Portions of the static trace generated by parsing example program's IR in Figure~\ref{fig:sa:running_example}.}
    \vspace{-3mm}
    \label{fig:sa:static_trace}
\end{figure}

\subsection{Trace Analyzer}
\label{subsubsec:sa:trace_analyzer}
The task of the trace analyzer can be divided into two steps. The steps are CFG generation and basic block count calculation.

\vspace{5pt}

\textit{\textbf{A) Control Flow Graph Generation :}}
We take the static trace as input and generate CFG of the target function in \emph{dot} format. The CFG provides the control flow and memory accesses in each basic block, probabilities of taking a branch in CFG, and basic block execution counts based on the program's input.

First, we calculate and enlist different types of instructions each basic block executes. Then, we identify the successors and predecessors of each basic block. To identify the successors, we look at the last instruction of the basic block. For example, if the last instruction is \emph{`br'} and has only one operand, then it is an unconditional branch, and the successor basic block is listed as its operand. If the number of operands is more than one, then it is a conditional branch, and the branch taken will depend on the compare instruction mentioned as the first operand of the \emph{`br'} instruction. We mark these basic blocks as the successors of the basic block ending with the conditional \emph{`br'} instruction. Similarly, we also identify all possible predecessors of a basic block. Please note that we also store the position of the first instruction of successor or predecessor basic blocks as the successor or predecessor's position in \emph{C} program which is used in the next step. At this point, we have gathered all the necessary information to construct the target function's basic block-level control flow graph.

\vspace{5pt}

\textit{\textbf{B) Getting Basic Block Execution Counts :}}
Once we have a function's basic block level control flow graph, we can measure the exact basic block execution count with the help of transition probabilities from one basic block to another. This transition probability is dependent on program input size. Let us consider a basic block $BB_j$ as a node in the CFG, which has predecessor basic blocks $BB_{ji_1}, BB_{ji_2}, ..... BB_{ji_m}$. Let us also assume that $BB_{jk_1}, BB_{jk_2}, ..... BB_{jk_n}$ are successor basic blocks of $BB_j$. Therefore, the predecessor and successor basic blocks of $BB_j$ satisfy the following linear balance equation.
    
\begin{equation}
\label{eq:sa:balance}
    \sum_{i \in Pred} ^{}P_{ij} \times N_i = \sum_{k \in Succ} ^{}P_{jk} \times N_k
\end{equation}
\vspace{6pt}

\noindent
where $P_{ij}$ is the transition probability from predecessor block $BB_i$ to $BB_j$, $P_{jk}$ is the transition probability from predecessor block $BB_j$ to $BB_k$, and $N_i$, $N_k$ are the execution counts of $BB_i$ and $BB_k$ respectively. Since the entry basic block of a kernel/program is executed once, $N_1$ is 1. Thus, for \emph{M} number of basic blocks in the CFG, we form \emph{M - 1} homogeneous linear equations and one non-homogeneous equation (for the entry basic block). We measure the transition probabilities from knowledge of code or using offline coverage tools. We leverage the location of instructions collected using the LLVM IR reader. It generates a \emph{lua} script where we fill up the transition probabilities. Figure~\ref{fig:sa:transition_prob} shows the generated script for our example program filled up with transition probabilities. As an example, in the Figure, the value of $T\_4\_5$ is found in line 4, column 5 of the input example program. These are the probabilities of taking a branch if the condition is false. Thus, the values of $T\_4\_5$, $T\_7\_9$, and $T\_10\_13$ are 1/101, 1/201, and 1/301, respectively. Consequently, we recursively solve equation~\ref{eq:sa:balance} for every basic block starting with the entry basic block in order to measure the execution count of all basic blocks. Finally, we can use these execution counts to measure the apriori probability of executing a basic block, $P(BB_i)$.

\begin{figure}[htbp]
    \centering
    \lstset{style=codeListStyle}
    \begin{lstlisting}[numbers=left]
local BranchingProbs = { --TODO: Fill in static values for the various branching probabilities for your instance
    T_11_13 = 1/301, -- main:9:[for.end]
    T_5_5 = 1/101, -- main:13:[for.end17]
    T_8_9 = 1/201, -- main:11:[for.end14]
}
return BranchingProbs
    \end{lstlisting}
    \vspace{-5mm}
    \caption{Generated Lua script to fill up with branch probabilities.}
    \vspace{-4mm}
    \label{fig:sa:transition_prob}
\end{figure}

\begin{figure}[htbp]
    \centering
    \includegraphics[width=0.47\textwidth]{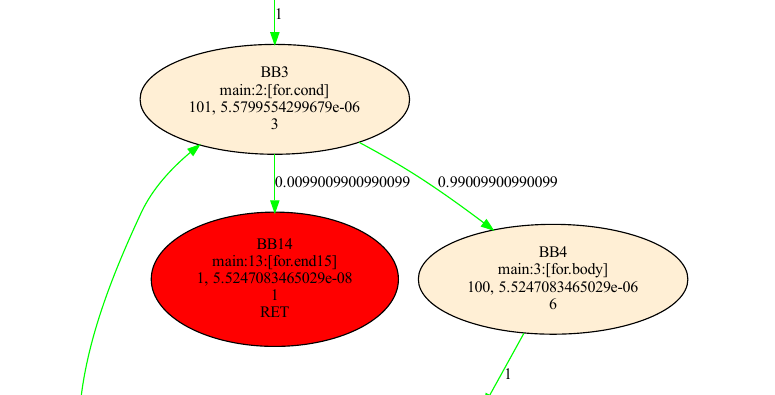}
    \vspace{-3mm}
    \caption{Snippet of our CFG showing nodes with execution count and edges with transition probabilities.}
    \vspace{-3mm}
    \label{fig:sa:cfg}
\end{figure}

\subsection{CFG Analyzer}
\label{subsubsec:sa:cfg_analyzer}
Once we have the CFG with basic block execution counts and branch probabilities, we make a virtual trace of the memory trace from it. We use the PygraphViz python library to read the CFG, stored in \emph{.dot} format. As shown in the example CFG in Figure~\ref{fig:sa:cfg}, each node represents a basic block of the target function, and the edges represent the direction from one node to another. The graph also shows the probabilities of taking a particular edge. We start traversing the CFG starting from the entry node. When we find multiple possible edges, we take the edge with the maximum probability count and mark it as a visited edge. Next time if we visit the same node, we take another unvisited edge with maximum probability. Thus, we find the path that has the maximum probability. Upon finding the path, we identify any loop within the path. If we find a loop, we annotate the path with loop symbols, determine the loop bounds, and \emph{loop control variable} collected in step~\ref{subsubsec:sa:bcread}. Thus, we find the loop annotated execution path from the CFG. The resultant path for our example program looks like the following.

\vspace{3pt}

\noindent
\textsf{\textbf{BB1 → BB2 → [100\textasciitilde i → BB3 → BB4 → [200\textasciitilde j → BB5 → BB6 → [300\textasciitilde k → BB7 → BB8 → BB9 → ] → BB7 → BB10 → BB11 → ] → BB5 → BB12 → BB13 → ] → BB3 → BB14}}

\vspace{3pt}

Replacing the basic blocks in the path with corresponding memory accesses of each basic block, we find the loop annotated static memory trace. The resultant trace for our running example program looks like the one below.

\vspace{3pt}

\noindent
\textsf{\textbf{retval → A → i → [100 → i → i → result → result → j → [200 → j → result → j → A → result → k → [300 → k → k → k → result → result → k → k → ] → k → result → i → j → arrayidx11\textasciitilde i-j → j → j → ] → j → i → i → ] → i}}

\subsection{Reuse Profile Calculation from Static Trace}
Finally, we deploy a recursive algorithm to efficiently calculate the reuse profile from this trace. Since this is a loop annotated memory trace, we leverage the annotation to calculate the reuse profile for the first two iterations of the loop and predict the reuse profile for the N number of iterations. When a program executes a loop for the first time, it sees the memory accesses that occur before the loop starts. For example, the innermost third loop is executed 300 times for every time program control goes to the loop. For the first iteration, it will see the address from the second-level loop. However, for the second to $N^{th}$ iteration, the second iteration's addresses will be repeated, and the loop will repeatedly see its own memory accesses as previous accesses ( \emph{k, k, k, result, k, k}). So, the reuse profile for the third to $N^{th}$ iteration of the loop is the same as that of the second iteration. Thus, by calculating the reuse profile of the second iteration, we can effectively calculate the reuse profile of the 3rd to $N^{th}$ iteration.
\begin{algorithm}[htbp]
    \begin{algorithmic}[1]
    \Procedure{$CalcReuseProfileRec$}{$idx$, $mem\_trace$}
        \State $reuse\_profile \gets \{\}$
        \While{$idx$ $<$ $len(mem\_trace)$}
            \State $addr \gets mem\_trace[idx]$
            \If{$addr[0]$ $==$ $`['$} \textcolor{codegreen}{\Comment{New loop starts}}
                \State $loop\_count \gets int(addr[1:])$
                \State $nxt\_idx, rp\_itr0 \gets CalcReuseProfileRec(idx + 1, mem\_trace)$ \textcolor{codegreen}{\Comment{Reuse profile of first loop iteration}} \label{first_iteration_call}
                \State $reuse\_profile \gets MergeRPs(reuse\_profile, rp\_itr0, 1)$
                \State $past\_trace \gets mem\_trace[:nxt\_idx]$
                \State $loop\_trace \gets mem\_trace[idx + 1 : nxt\_idx]$
                \State $\_, rp\_itr\_otr \gets CalcReuseProfileRec(nxt\_idx, past\_trace + loop\_trace)$ \textcolor{codegreen}{\Comment{Reuse profile of other loop iterations}}
                \State $reuse\_profile \gets MergeRPs(reuse\_profile, rp\_itr\_otr, loop\_count - 1)$
                \State $idx \gets nxt\_idx$ \textcolor{codegreen}{\Comment{Updates idx to end of loop}} 
            \ElsIf{$addr$ $==$ $`]'$}
                \State \Return $idx, reuse\_profile$
            \Else
                \State $window \gets list(filter(lambda$ $ x : x[0] \neq `['$ $\land$ $x[0] \neq `]', mem\_trace[:idx]))$
                \State $dict\_sd \gets \{\}$
                \State $addr\_found \gets False$
                \For{$w\_idx$ $\textbf{in}$ $range(0, len(window))$}
                    \State $w\_addr \gets window[-w\_idx -1]$
                    \If{$w\_addr == addr$}
                        \State $addr\_found \gets True$
                        \State $\textbf{break}$
                    \EndIf
                    \State $dict\_sd[w\_addr] \gets True$
                \EndFor
                \If{$addr\_found$}
                    \State $r\_dist = len(dict\_sd)$
                    \State $reuse\_profile[r\_dist]$ $\pluseq$ $1$
                \Else
                    \State $reuse\_profile[-1]$ $\pluseq$ $1$
                \EndIf
            \EndIf
            \State $idx \pluseq 1$
        \EndWhile
        \State \Return $idx, reuse\_profile$
    \EndProcedure
    \end{algorithmic}
    \caption{Reuse Profile Calculation from Static Trace}
    \label{alg:sa:rd_recursive}
\end{algorithm}


Algorithm~\ref{alg:sa:rd_recursive} shows how we recursively calculate the reuse profile from the static trace. \emph{CalcReuseProfileRec} function is called with idx=0 and the static trace as input. Then, it starts traversing the trace. Whenever it finds an entry that is not a loop-bound notation (\emph{`[' or `]'}), it calculates the reuse distance for that entry. If it encounters a loop start notation (addr[0] == `['), first it extracts loop iteration count (\emph{loop\_count}). Then \emph{CalcReuseProfileRec} is recursively called after increasing \emph{idx} by one (line 7). The resultant reuse profile from this call denotes the reuse profile for the first iteration of the loop. It is then accumulated with the already calculated reuse profile of previous addresses that appear in the trace before the loop. The recursive call also returns the index (\emph{nxt\_idx}) in the trace where the loop ends (when `]' is encountered). Then, we identify the portion of the trace under the loop. For the second to $N^{th}$ iteration, the loop repeatedly sees its memory addresses as previous accesses. So, we add this loop trace to the previous trace, including the loop trace, and calculate the reuse profile for the second iteration of the loop. Then, this reuse profile is multiplied by \emph{loop\_count - 1} to get the reuse profile of the second to $N^{th}$ iteration.

When memory access is due to an array reference, we calculate its reuse distance differently. We determine if the array is being accessed from inside loops. If the loop control variables and array indexing variables are the same, then we consider that a new memory location will be referenced each time. Otherwise, the same memory location will be accessed in every iteration. We also determine if the array index is always constant (\emph{e.g.} array[i][5]) or not and calculate the reuse profile for that case differently. Further details of our array reference processing approach are beyond the paper's scope.

\begin{figure}[htbp]
\centering
\subfigure[{Memory transactions in bytes}]{
\label{fig:sa:membytes}%
\includegraphics[width=0.45\linewidth]{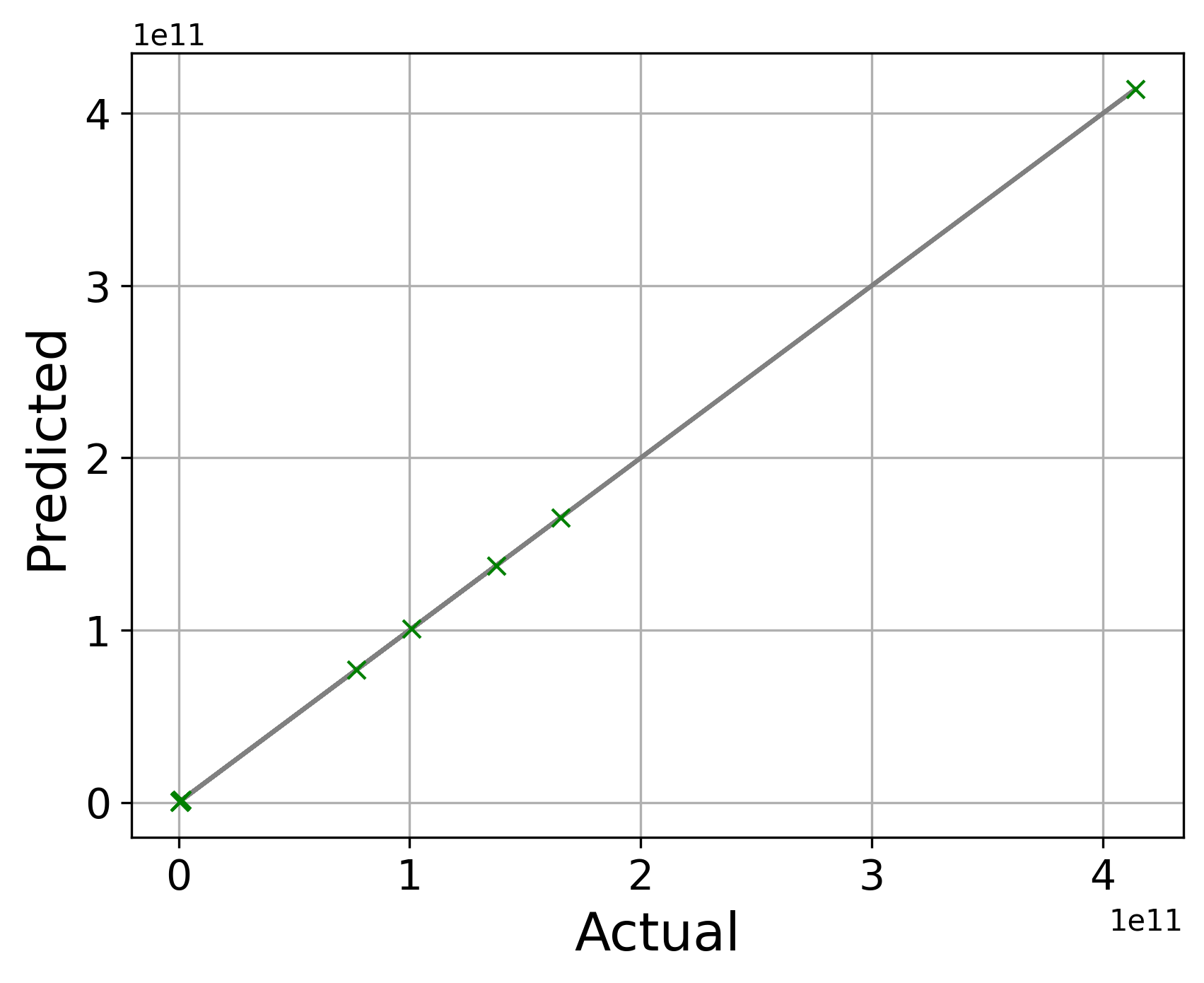}}%
\qquad
\subfigure[{Number of load operations}]{
\label{fig:sa:load}%
\includegraphics[width=.44\linewidth]{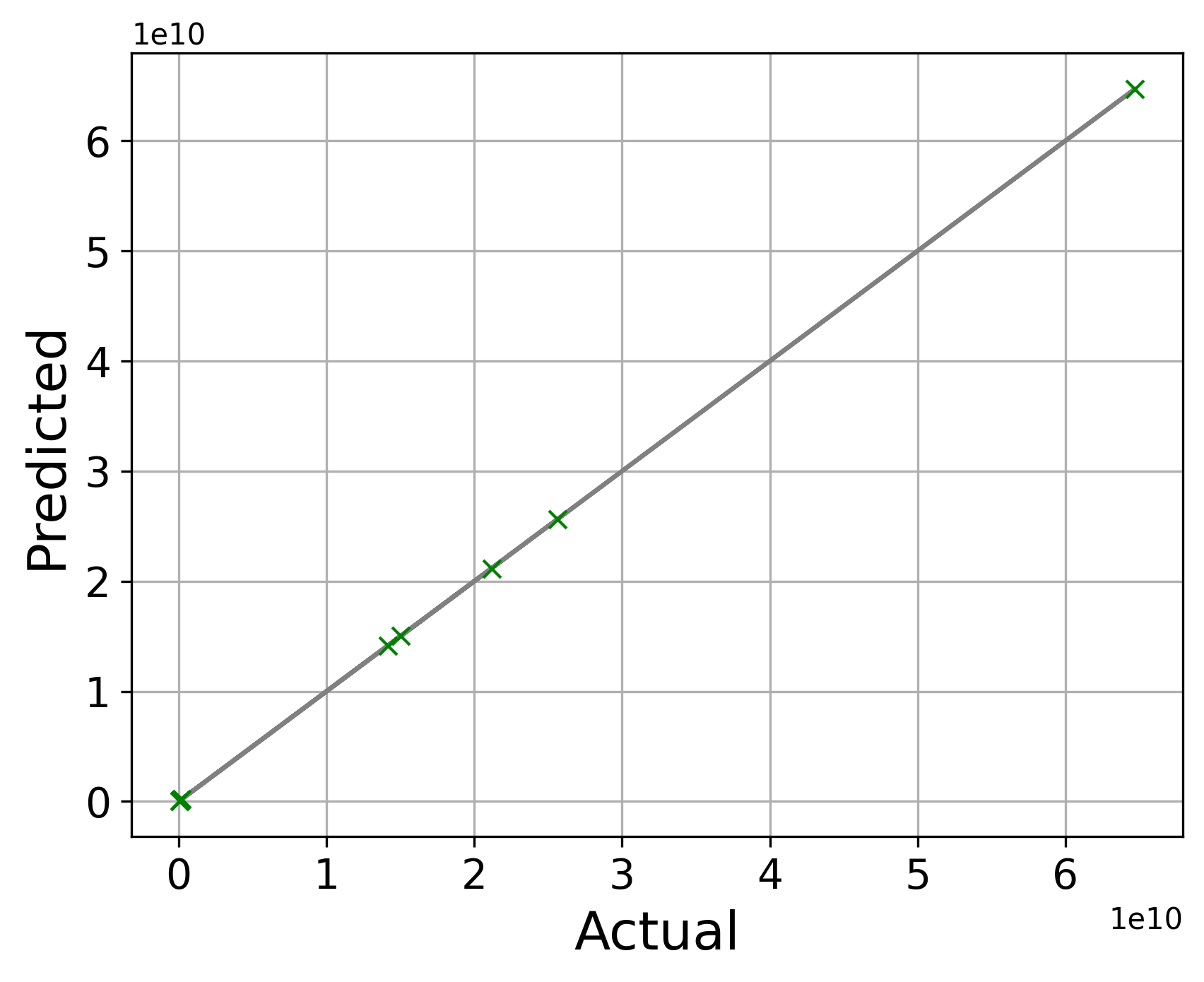}}%
\qquad
\subfigure[{Number of store operations}]{
\label{fig:sa:store}%
\includegraphics[width=.44\linewidth]{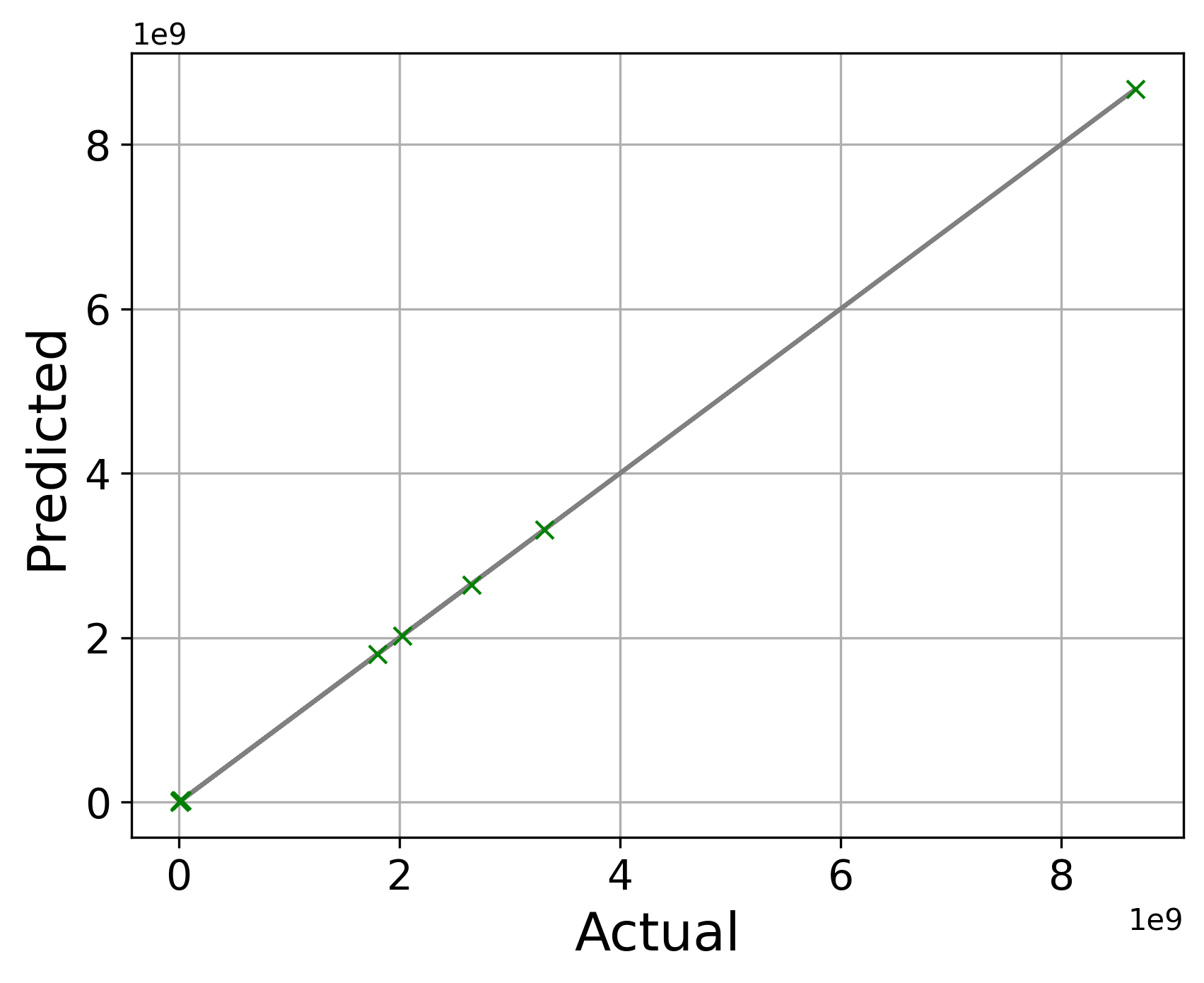}}%
\qquad
\subfigure[{Number of unconditional branch operations}]{
\label{fig:sa:uncond-br}%
\includegraphics[width=.44\linewidth]{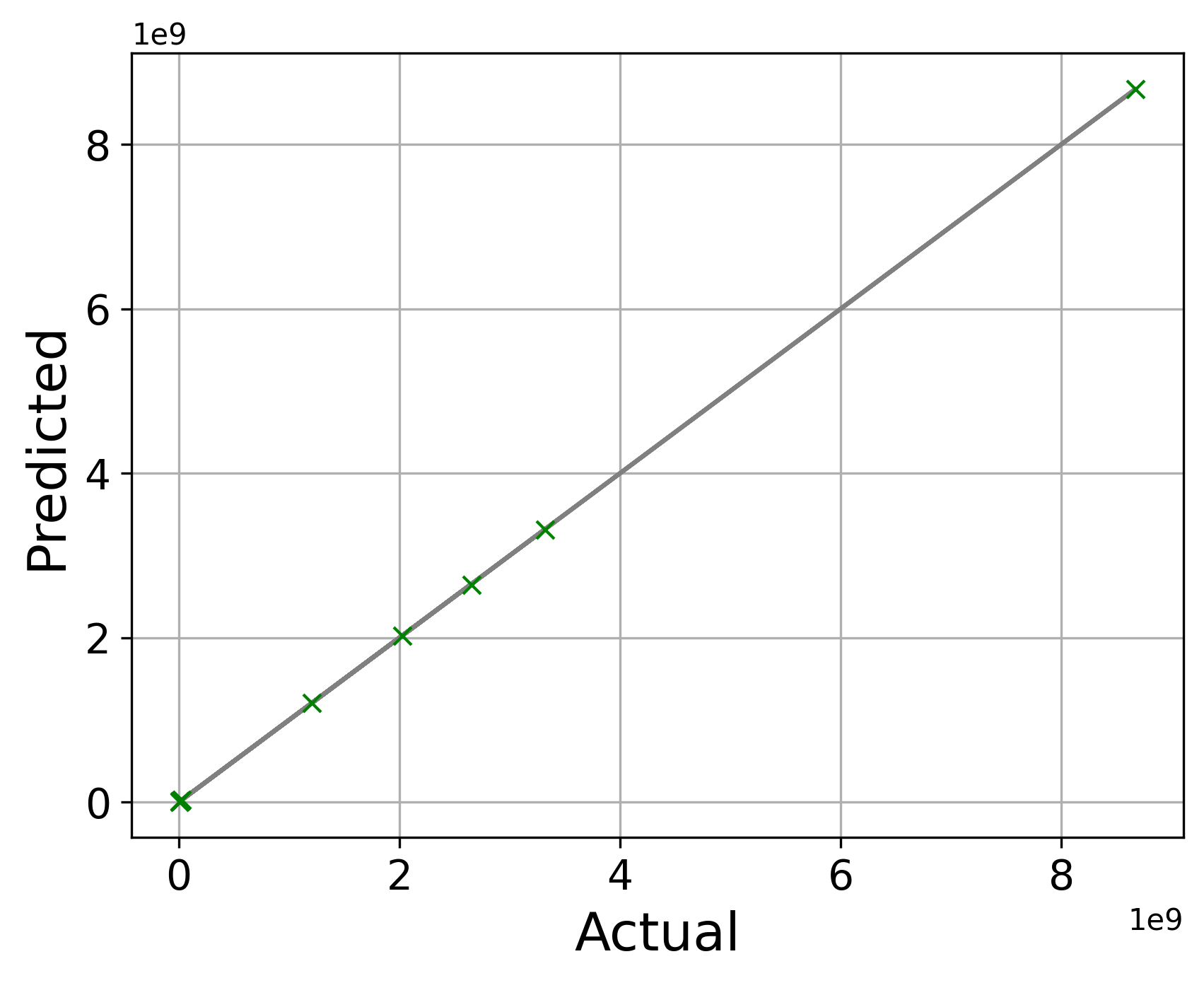}}%
\qquad
\subfigure[{Number of conditional branch operations}]{
\label{fig:sa:cond-br}%
\includegraphics[width=.44\linewidth]{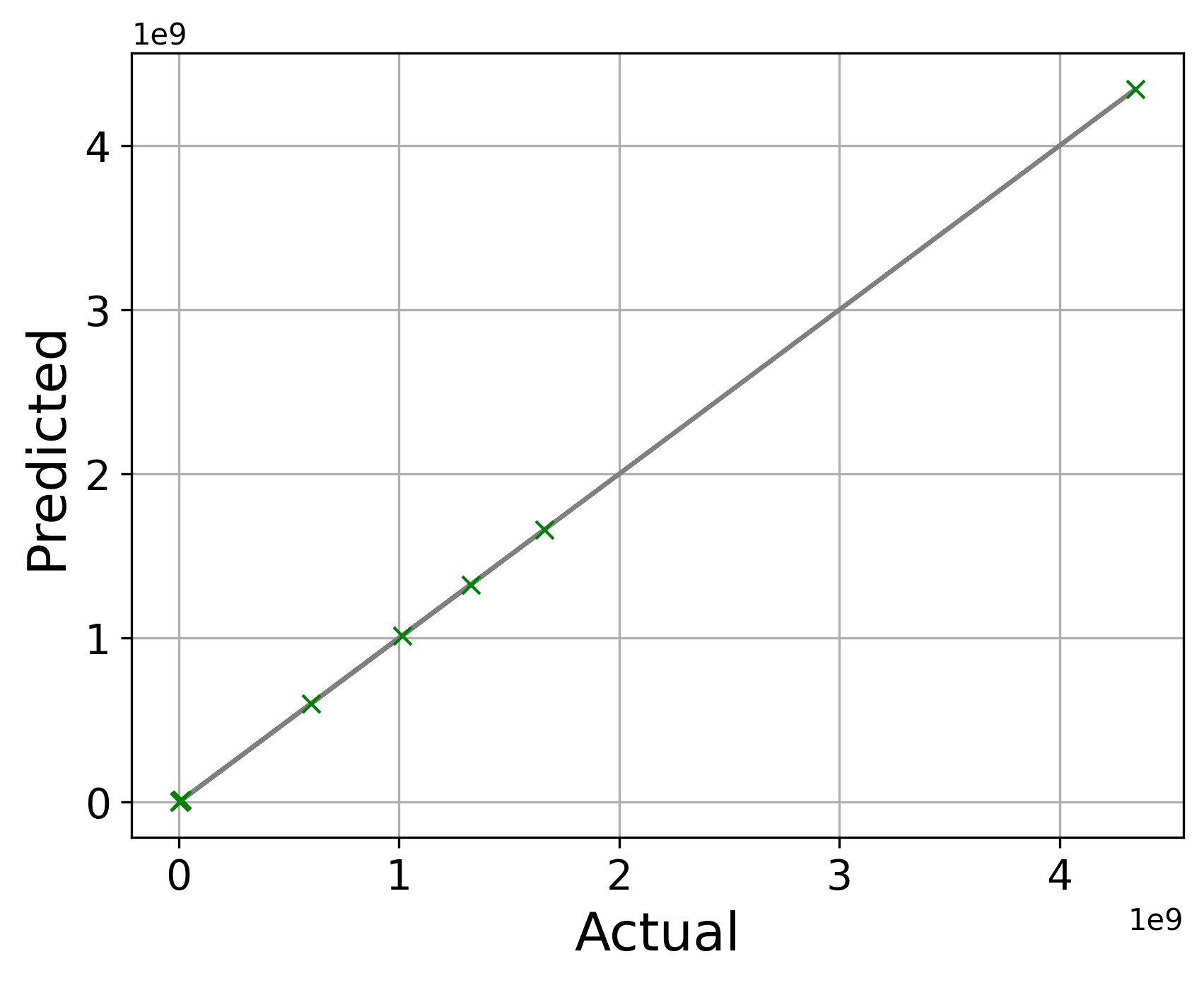}}%
\qquad
\subfigure[{Number of floating point operations}]{
\label{fig:sa:flops}%
\includegraphics[width=.44\linewidth]{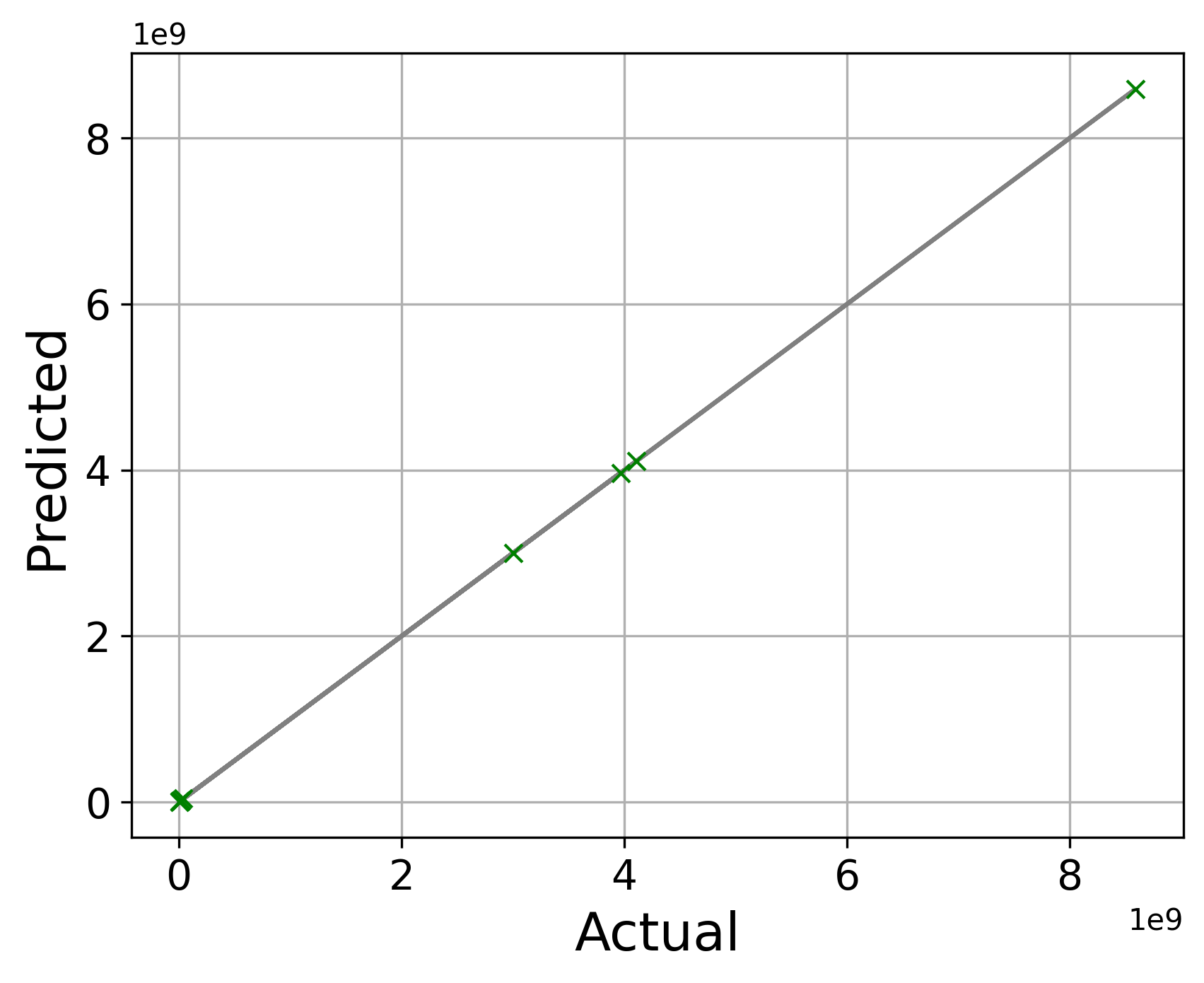}}%
\qquad
\caption{Comparison of different performance metrics between Byfl and static analyzer. Each green 'x' mark denotes the predicted value, whereas the gray line denotes the actual result from Byfl.}
\label{fig:sa:results-metrices}%
\end{figure}

%% file: sections/results.tex
In this section, we validate the results of the static analyzer. First, we validate the programs' performance metrics using ten applications from different domains from the Polybench~\cite{polybench} benchmark suite. The applications are listed in Table~\ref{table:sa:benchmarks}. We choose the large dataset of the applications in the Polybench benchmark suite. We compare our results with the results from the Byfl~\cite{Byfl}. Byfl is an LLVM-based hardware-independent dynamic instrumentation tool for application characterization, which makes it a suitable choice to validate our results.

Figure~\ref{fig:sa:results-metrices} shows the comparison of different metrics between the static analyzer and Byfl, where Figures~\ref{fig:sa:membytes}, ~\ref{fig:sa:load}, ~\ref{fig:sa:store}, ~\ref{fig:sa:uncond-br}, ~\ref{fig:sa:cond-br}, and~\ref{fig:sa:flops} show the comparison of total memory transactions, load operations, store operations, unconditional branches, conditional branches, and floating point operations respectively for the benchmark kernels. In the figures, the gray line denotes the result from Byfl, where each green 'x' mark denotes the predicted value from the static analyzer tool. Our results match with Byfl accurately, given that the same LLVM version is used to compile Byfl and in our tool.

\begin{figure}[htbp]
    \centering
    \vspace{-3mm}
    \includegraphics[trim=25mm 20mm 30mm 30mm, clip, width=0.45\textwidth]{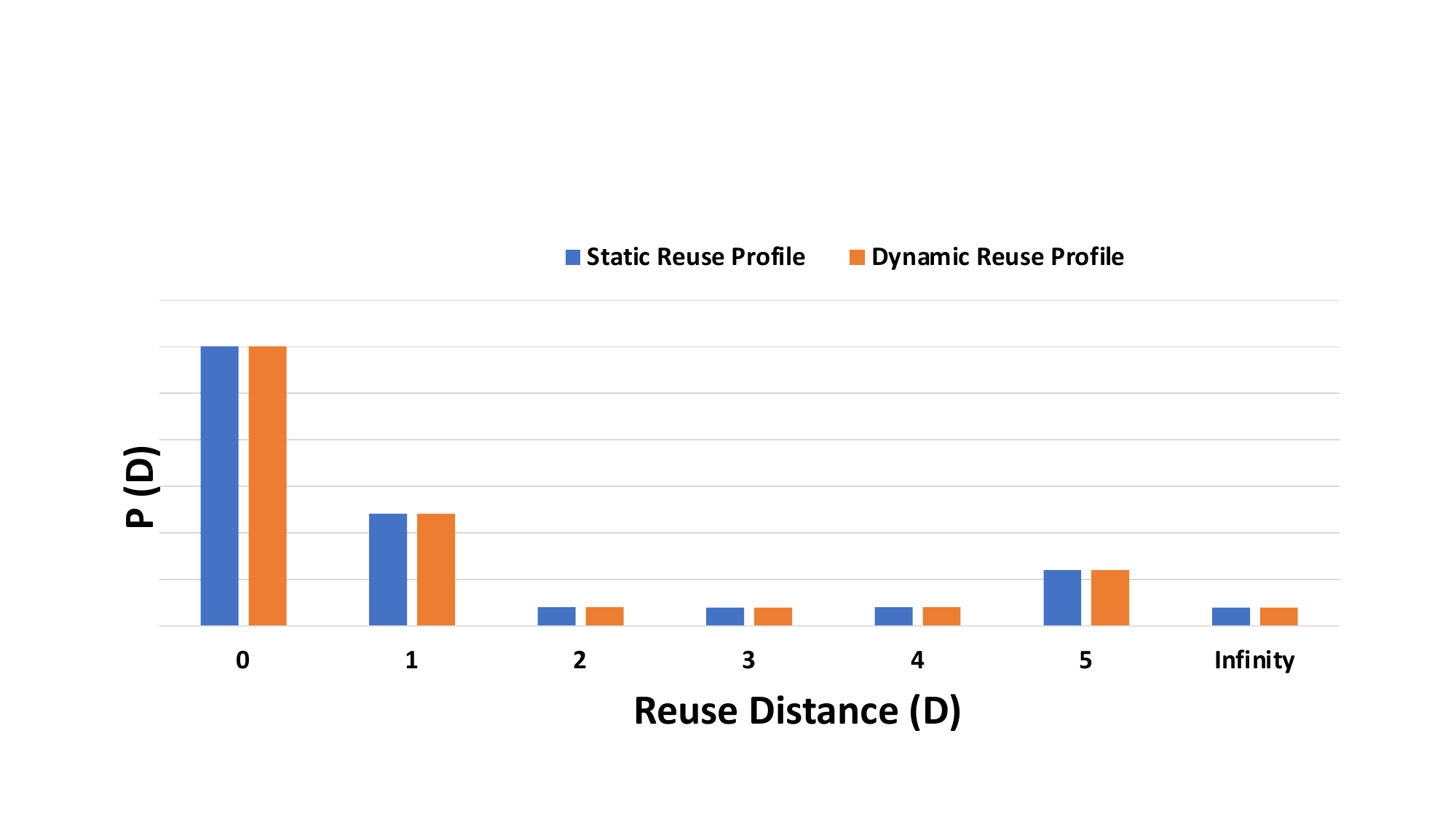}
    \vspace{-3mm}
    \caption{Comparison between statically and dynamically calculated reuse profiles.}
    \label{fig:sa:result-rd}
    \vspace{-3mm}
\end{figure}

\textbf{Reuse Profile Validation:}
In this section, we validate our static reuse distance calculation method. To validate, we instrument our target kernel using Byfl~\cite{Byfl} and generate a dynamic memory trace. Later, using the parallel tree-based reuse distance calculator, Parda~\cite{PARDA:Niu}, we calculate the dynamic reuse profile of the program. We compare this dynamic reuse distance with the statically calculated reuse profile. Figure~\ref{fig:sa:result-rd} shows the reuse profile for the synthetic benchmark in Figure~\ref{fig:sa:running_example}, where the X axis represents reuse distance and the Y axis represents the probability of a particular reuse distance P(D). Results show that our approach can calculate reuse profiles correctly with $100\%$ accuracy.

%% file: sections/limitations.tex
There are several limitations to our approach. First, our reuse calculation approach still needs to be improved in handling pointer/array references. As a result, we still cannot accurately calculate the reuse profiles for the applications listed in Table~\ref{table:sa:benchmarks}. To support complex cases array references, we must further modify the algorithm~\ref{alg:sa:rd_recursive} to consider complex array reference cases.

Another limitation of our approach is that calculating branch probabilities for data-dependent branch cases can be difficult. For example, each iteration may vary the branch probability for an inner loop's condition. Although we can calculate overall branch probabilities using arithmetic summation to handle this issue, it should be fully automated. In addition, our static analyzer has yet to support multiple functions.

%% file: sections/conclusion.tex
Static analysis can play a vital role in performance modeling and prediction. 
 This paper introduces an architecture-independent LLVM-based static analysis approach to estimate a program's performance characteristics, including the number of instructions, total memory operations, etc. It also presents a novel static analysis approach to predict the reuse profile of the application kernel, which can be used to predict cache hit rates. The prediction time of our approach is invariant to the input size of the program, which makes our prediction time almost constant and, thus, highly scalable. The results show we can statically predict reuse distance profiles of specific applications quickly and accurately compared to binary instrumentation approaches. Thus, our approach is highly suitable for early-stage design space exploration.